\documentstyle[preprint,aps,psfig]{revtex}
\begin{document}
\preprint{}
\title{
Effect of Finite Impurity Mass on the Anderson Orthogonality Catastrophe
in One Dimension}
\author{H. Castella}
\address{ Institut Romand de Recherche Num\'erique en Physique des
Mat\'eriaux (IRRMA), \\
PHB-Ecublens, CH-1015 Lausanne, Switzerland\\
and\\
Department of Physics, Ohio State University,\\ 
174 West 18th Ave., Columbus, OH, 43210-1106.~~~\cite{adres}}
\date{Received\ \ \ \ \ \ \ \ \ \ \ }
\bigskip\bigskip
\maketitle

\begin{abstract}
A one-dimensional tight-binding Hamiltonian describes the evolution of a 
single impurity interacting locally with $N$ electrons.
The impurity spectral function has a 
power-law singularity $A(\omega)\propto\mid\omega-\omega_0\mid^{-1+\beta}$
with the same exponent $\beta$ that characterizes the logarithmic decay of 
the quasiparticle weight $Z$ with 
the number of electrons $N$, $Z\propto N^{-\beta}$. The exponent $\beta$ is 
computed by (1) perturbation theory in the interaction strength and
(2) numerical evaluations with exact results for small systems and 
variational results for larger systems. A nonanalytical behavior of $\beta$ is  
observed in the limit of infinite impurity mass. For large interaction 
strength, the exponent depends strongly on the mass of the impurity in 
contrast to the perturbative result. 
\end{abstract}

\pacs{PACS numbers: 71.27.+a, 72.10-d}

\section{Introduction}

Anderson studied the effect of a static impurity potential on conduction 
electrons in metals \cite{and} and showed that the groundstate of the 
electrons is strongly renormalized by the local potential of the impurity 
and has an overlap with the unperturbed state, or quasiparticle weight $Z$, 
vanishing as $Z\propto N^{-\beta}$ with increasing number of electrons $N$.
This effect, known as the Anderson orthogonality catastrophe, has its origin 
in an infrared singularity due to shake-up processes of the 
electron sea in the presence of the impurity potential \cite{hop} and 
signals the failure of the quasiparticle picture to describe the 
low-energy excitations. 

The infrared singularity also affects the optical properties of metals. 
The core-level hole  created by an X-ray disturbs the conduction electrons 
similarly to an impurity potential.
On one hand, the X-ray photoemission spectrum is asymmetrically 
broadened above the threshold \cite{wer,don}. On the other hand, the 
X-ray absorption spectrum has a strongly enhanced threshold, 
the so-called Fermi-edge singularity \cite{mah}. 

These singularities in the optical spectra apply for a static core hole,
i.e., an infinite-mass hole (or impurity). For a finite-mass hole and an 
isotropic band dispersion, the infrared singularity does not occur 
in three dimensions because the hole recoil strongly restricts the 
number of low-energy excitations \cite{mha,gav}; as a consequence, the edge 
singularities disappear. In one dimension, however, the infrared singularity 
persists even for a finite-mass hole \cite{kop}. 
The observation of an enhanced threshold in UV-absorption 
spectra of doped semiconductor quantum wires was interpreted 
as a Fermi-edge singularity \cite{cal} and stimulated renewed interest 
in the one-dimensional problem \cite{mul,per}. 

The present work studies 
the Anderson orthogonality for a finite-mass impurity in a simple 
one-dimensional model and focusses on the impurity recoil.
Although the infrared singularity occurs for both an 
infinite- and a finite-mass impurity in one dimension, the recoil 
plays an important role on the critical exponent $\beta$ which does not 
extrapolate to the static-impurity value in the infinite-mass limit.
The dependence of the exponent on the impurity mass is investigated 
analytically by perturbation theory and numerically by a variational approach 
in the nonperturbative regime. The results are eventually compared to 
analytical calculations from Ref. \cite{thi} which studied the heavy-mass 
and strong-coupling regime using a path-integral formalism \cite{guinea}. 

The next section presents the model and the results. In section \ref{pertu}, 
a perturbation analysis of the impurity spectral function is performed along 
the line of Ref. \cite{kop}. Section \ref{numeri} calculates numerically 
the critical exponent using a variational approach. 

\section{Model and results}
\label{model}
This section presents the model and summarizes the known results on the 
impurity spectral function and the Anderson orthogonality catastrophe.
At the end of the section, the main results of this work are briefly 
described.

The model describes a single impurity and $N$ spinless electrons moving 
on a chain of $L$ sites with periodic boundary conditions. Within a 
tight-binding approximation with nearest-neighbor hopping, the band energies 
are $-2t_h\cos k$ and $-2t\cos k$ for the impurity and the 
electrons, respectively. Further, the impurity and the electrons 
feel an interaction $U$ when they sit on the same site. 
Although this study is restricted to a single impurity, it is convenient to 
write the Hamiltonian $\hat{H}$ in second-quantized form with
creation operators $c_i^{\dagger}$ for an electron on site $i$, and 
$d_i^{\dagger}$ for an impurity:
\begin{equation}
\hat{H}=-t\sum_{i=1}^L \left( c^{\dagger}_i c_{i+1} + h.c.\right) 
    - t_h \sum_{i=1}^L \left( d^{\dagger}_i d_{i+1} + h.c. \right)
    + U \sum_{i=1}^L d^{\dagger}_i d_i c^{\dagger}_i c_i.
\label{H}
\end{equation}
The interaction is attractive in order to describe a hole in a valence band. 
For this particular model, however, the repulsive and attractive cases are 
related by a particle-hole transformation for the electrons 
$\tilde{c_j}= (-1)^jc_j^{\dagger}$. In the rest of the paper all the 
results are presented for the repulsive case; the corresponding results for 
the attractive interaction are obtained by the transformation 
$\rho\to 1-\rho$ where $\rho=N/L$ is the density of conduction electrons.

The spectral function of the hole (or impurity) $A(q,\omega)$ 
describes the photoemission response within the sudden approximation, i.e.,
neglecting interaction with the outgoing electron \cite{alm}. It has 
a spectral decomposition in terms of eigenstates, $\mid\psi_n\rangle$, and 
eigenenergies, $E_n$,  of the Hamiltonian (\ref{H}) in presence of the 
impurity, and of the groundstate wavefunction, $\mid\phi_0\rangle$, and 
energy, $\tilde{E}_0$, in absence of impurity:
\begin{equation}
A(q,\omega)=\frac{1}{\pi}\hbox{Im}G_q(\omega)=\sum_{n}\mid\langle\psi_n\mid
d_q^{\dagger}\mid\phi_0\rangle\mid^2\delta(\omega-E_n+\tilde{E}_0).
\label{Aq}
\end{equation}
In the spectral decomposition of $A(q=0,\omega)$, the spectral weight of the 
groundstate $Z=\mid\langle\psi_0\mid d_{q=0}^{\dagger} 
\mid\phi_0\rangle\mid^2$ may remain finite in the thermodynamic limit, giving 
rise to a quasiparticle peak in the spectral function. This is the usual 
situation when the quasiparticle picture applies. 

{\it Static impurity} ($t_h=0$). 
The interaction causes the quasiparticle picture to break down \cite{and}: 
the spectral weight scales to zero with increasing number of fermions $N$
as $Z\propto N^{-\beta}$, the groundstate of the interacting
system being orthogonal to the quasiparticle state $d^{\dagger}\mid
\phi_0\rangle$ for $N\to\infty$. This is  known as the Anderson orthogonality 
catastrophe. The exponent is related to the phaseshift $\delta_F$ of an 
electron at the Fermi energy scattered off the static impurity:
\begin{equation}
\beta(t_h=0)=(\delta_F/\pi)^2.
\label{beta0}
\end{equation}
For the present model, the phaseshift depends on $U$ and on the density of 
states at the Fermi energy $N_F=1/(2\pi\sin k_F)$:  $\delta_F=
-\arctan\left(\pi U N_F\right)$. Furthermore, the spectral function has, 
instead of a quasiparticle peak, a power-law singularity at the threshold 
$\omega_0=E_0-\tilde{E}_0$ with the critical exponent $1-\beta(t_h=0)$ 
\cite{don}:
\begin{equation}
A(\omega)\propto\frac{1}{\mid\omega-\omega_0\mid^{1-\beta}}.
\end{equation}
This singularity in $A(\omega)$ is observed in X-ray photoemission of 
metals \cite{wer}. 

{\it Finite mass impurity} ($t_h>0$). While an infinite-mass impurity acts 
as an external potential on the electrons, the impurity recoil further 
complicates the many-body problem.
Despite this complexity, the eigenenergies and eigenstates of $\hat{H}$ 
are known exactly for the special case $t_h=t$ \cite{mcg}. 
Using the exact solution, the spectral function for $q\simeq 0$ is computed 
in Ref. \cite{cas}. It has no quasiparticle peak because of the Anderson 
orthogonality catastrophe and has a power-law singularity with an exponent:
\begin{equation} 
\beta(t_h=t)=2\left(\delta'_F/\pi\right)^2
\label{BAz}
\end{equation}
The exponent is given by the phaseshift of a single electron at $k_F$ 
scattering off a finite-mass impurity $\delta'_F=-\arctan\left(\pi 
U N_F/2 \right)$. Notice the similarity in the exponents for 
$t_h=t$ and $t_h=0$, which are both expressed in terms of phaseshifts. 
The phaseshifts, however, differ since the number of states 
contributing to the Anderson orthogonality is reduced from $UN_F$ to 
$UN_F/2$ between a static and finite mass impurity, respectively. The 
origin of this difference is discussed in Section \ref{pertu} for the 
perturbative results. 

The present work presents calculations of the exponent $\beta(t_h)$ for 
different hopping parameters $t_h$, interaction strengths $U$, and 
electron densities $\rho$. The exponent $\beta(t_h)$ is computed (1) 
analytically using perturbation theory in $U$ and (2) numerically in the 
nonperturbative regime using a variational wavefunction proposed by 
Edwards \cite{edw}. The exponent is extracted numerically from finite 
size results using a precise scaling law for $Z$ as a function of $N$ and a 
numerical fit of the data as $N\to\infty$. The main results of this study 
are summarized now. Sections \ref{pertu} and \ref{numeri} will give a 
detailed description of the perturbation calculations and the 
numerical simulations, respectively. 

{\it Mass dependence of the exponent.}
The perturbative results indicate that the exponent $\beta$, to order 
$(U/t)^2$, is independent of $t_h$ for $t_h>0$ and equals $(UN_F)^2/2$. 
For a finite $U$, however, the exponent $\beta(t_h)$ calculated numerically 
does depend on $t_h$ and its dependence increases with increasing $U$, 
as illustrated in Fig. \ref{one} that shows the exponent normalized to 
its value for $t_h=t$, $\beta(t_h)/\beta(t_h=t)$, as a function of $t_h$. In 
the strong coupling limit, $U=\infty$, $\beta$ varies quasi linearly with $t_h$.

{\it Discontinuous exponent in the heavy-mass limit.}
The perturbative calculations predict a discontinuous exponent in the 
limit of a flat impurity dispersion:  
$\lim_{t_h\rightarrow 0} \beta(t_h) = \beta(t_h=0)/2$. This discontinuity 
is due to the irrelevance of backscattering processes for $t_h>0$  
because of the finite recoil energy involved. The numerical results in 
Fig. \ref{one} illustrate the nonanalyticity for a finite $U$:  
when $t_h\rightarrow 0$ the exponent does not extrapolate to the static value 
which is indicated by the filled symbols at $t_h=0$. 
The numerical results can be compared to calculations of A. Rosch and T. Kopp 
\cite{thi} for the exponent in the heavy-mass and strong-coupling regime. 
Their analysis, based on an effective action for the long-time behavior of 
the impurity propagator, also predicts a discontinuity 
$\lim_{t_h\rightarrow 0} \beta(t_h)/\beta(t) = \alpha\neq\beta(0)/
\beta(t)$ and gives at half filling $\alpha=1/4$ while at third filling
$\alpha=19/56\simeq 0.339$. The numerical results in Fig. \ref{one} 
suggest $\alpha\simeq 0.25$ for both half and third filling. The result 
at half filling is therefore in good agreement with their prediction while 
at third filling the value of $\alpha$ is significantly smaller.
Note however that the present work relies on a variational approach.

{\it Crossover behavior in the heavy-mass limit.}
A detailed analysis of the perturbative results for $t_h\ll t$ reveals 
a crossover in the scaling behavior of $\ln Z$ as a function of $\ln N$, as 
shown in Fig. \ref{two}. While $\ln Z$ closely follows the static-impurity 
behavior with a slope $\beta(t_h=0)$ for a number of fermions smaller than a 
crossover value $N_c$, it adjusts to the finite-mass behavior only 
for $N>N_c$. Further $N_c$ diverges as $t/t_h$ for $t_h\to 0$. Therefore, the 
discontinuity is only an asymptotic result, valid in the limit $N\to\infty$.

\section{Perturbation theory}
\label{pertu}

This section evaluates the impurity spectral function and spectral 
weight $Z$ perturbatively in $U$ and for an arbitrary hopping parameter 
$t_h$ in one dimension,  following Ref. \cite{kop} that computes 
$A(q=0,\omega)$ for the equal-masses case ($t_h=t$) in 
connection with the stability of the ferromagnetic state in the Hubbard model. 

{\it Spectral weight.} In the perturbative expansion, the first terms that 
renormalize the ground-state wavefunction correspond to the creation of  
a single particle-hole pair within the Fermi sea by an impurity of 
momentum $q=0$. The excitation energy is: $\Delta\epsilon(k_1,k_2)=
2t\cos k_1-2t\cos k_2+2t_h-2t_h\cos(k_1-k_2)$. The spectral weight has a 
cumulant expansion \cite{mha} that, up to second order in $U$, involves only 
these excitations:  
\begin{equation}
\ln Z=-\left(\frac{U}{L}\right)^2\sum_{k_1,k_2}
\frac{\Theta(k_F-|k_1|)\Theta(|k_2|-k_F)}{\Delta\epsilon(k_1,k_2)^2}.
\label{Zper}
\end{equation}
The sum over $k_1,k_2$ diverges logarithmically with increasing number 
of electrons $N$. As shown in the Appendix, a large $N$ expansion gives~: 
\begin{equation}
\ln Z=-\beta(t_h)\ln N+\alpha_0(t_h)+\alpha_1(t_h)/N+O(1/N^2).
\label{Zlaw}
\end{equation}
The logarithmic term dominates for large $N$ and gives rise to the Anderson 
orthogonality. The finite-size corrections are used in Section
\ref{numeri} for the numerical study of $\beta$.

The main result of this section is the evaluation of $\beta(t_h)$: 
\begin{eqnarray}
 \beta(0)=&(UN_F)^2  &\quad {\rm for}~t_h=0 
\nonumber\\
 \beta(t_h)=&\frac{1}{2}(UN_F)^2  &\quad {\rm for}~t_h>0
\label{beta}
\end{eqnarray} 
The exponent $\beta(t_h)$ is independent of $t_h$ as far as $t_h>0$. 
The infrared singularity is caused by forward-scattering 
processes with small momentum transfer $|k_2-k_1|\ll k_F$, 
which are gapless excitations for any $t_h$. The hopping $t_h$ is irrelevant 
since the impurity recoil energy $2t_h(\cos(k_1-k_2)-1)$ is negligible as 
compared to the particle-hole energy $2t(\cos k_2-\cos k_1)$. 

Further, the exponent has a discontinuity in the heavy-mass limit 
$\lim_{t_h\to 0}\beta(t_h)=\beta(t_h=0)/2$. The difference between the 
infinite- and the finite-mass exponents is simply related to scattering of 
one electron from one side of the Fermi surface to the other.
These so-called backscattering processes, which involve
a large momentum transfer $|k_2-k_1|\simeq 2k_F$, do not contribute to the 
infrared divergence for $t_h>0$ since the impurity recoil opens a gap 
$2t_h(\cos 2k_F-1)$. For $t_h=0$ however, both backscattering and 
forward-scattering processes are gapless. The number 
of low-energy excitations contributing to the infrared singularity is thus 
reduced by a factor of $2$ for a finite-mass impurity as compared to its value 
for the static impurity. 

The discontinuity in the exponent is an asymptotic result valid only for 
$N\to\infty$. For a finite number of fermions and a large but finite mass 
$0<t_h\ll t$ however, $\ln Z$ has a crossover as a function of $\ln N$, 
illustrated in Fig. \ref{two} where the spectral weight is computed 
numerically from (\ref{Zper}). The logarithm of the spectral weight has 
the slope $\beta(t_h)$ only for a number of electrons larger than a 
crossover value $N_c$, while for a small $N$ it follows the static-impurity 
behavior with a slope $\beta(t_h=0)$. The dashed lines indicate the asymptotic 
behaviors $-\ln Z=\beta(t_h)\ln N-\alpha_0(t_h)$, with the analytical 
$\beta$ from ({\ref{beta}) and $\alpha_0$ fitted to the value of 
$\ln Z$ for the largest size. The intercept of the asymptotes gives 
the crossover size $N_c$, which agrees very well with the estimate 
$N_c\simeq 0.3244t/t_h$, presented in the Appendix. Notice that $N_c$ diverges 
as $t/t_h$ for $t_h\to 0$ and the asymptotic regime is reached for a larger 
number of electrons the lower $t_h$. 
 
{\it Spectral function.} The spectral function is computed only at $q=0$ 
where it has a power law singularity. The propagator is:
\begin{equation}
G_{q=0}(\tau)=~-i~\langle\phi_0\mid d_{q=0}\exp(-i\hat{H}\tau)d_{q=0}^{\dagger}
\mid\phi_0\rangle~ e^{i\tilde{E}_0\tau}\Theta(\tau).
\end{equation}
The propagator has also a cumulant expansion and is 
written in terms of the density of particle-hole excitations 
$S(\omega)$ and a renormalized impurity energy $\tilde{\epsilon}_0$: 
\begin{equation}
G_{q=0}(\tau)=~-i~\exp\left[-i\tilde{\epsilon}_0\tau-U^2\int_0^{\infty}
S(\omega)
 \frac{1-\exp(-i\omega\tau)}{\omega^2}d\omega\right].
\label{cum}
\end{equation}
\[
S(\omega)=\frac{1}{\left(2\pi\right)^2} \int_{-\pi}^{\pi}\int_{-\pi}^{\pi}
\Theta(|k_1|-k_f)\Theta(k_f-|k_2|)\delta(\omega-\Delta\epsilon(k_1,k_2))
dk_1 dk_2.\]

The density $S$ has a linear frequency dependence for small 
$\omega$: $S(\omega)=(\beta(t_h)/U^2) \omega$ where $\beta(t_h)$ is the 
exponent of the Anderson orthogonality in (\ref{beta}). This linear 
behavior determines the low-frequency spectral function which has a 
power-law singularity at threshold with the exponent $1-\beta$ \cite{don}:
\begin{equation}
A(q=0,\omega)=\frac{\sin(\pi\beta)\Gamma(1-\beta)}{\pi
(\omega-\tilde{\epsilon}_0)^{1-\beta}}\Theta(\omega-\tilde{\epsilon}_0).
\end{equation}

This is however only an asymptotic result for frequencies smaller than 
a cutoff $W$. For $t_h=0$, the cutoff is of the order of the Fermi 
energy: $W\simeq 2t(1-\cos k_F)$. For $t_h>0$ however, the linear behavior 
of $S$ holds only for frequencies smaller than the impurity recoil energy 
and the cutoff is given by $W\simeq\min(2t(1-\cos k_F),2t_h(1-\cos 2k_F))$. 
For a heavy impurity the cutoff is of the order of the impurity recoil energy 
$t_h(1-\cos 2k_F)$ rather than the Fermi energy and the asymptotic result is 
valid only in a very narrow frequency range. Furthermore, the density of 
excitations exhibits a crossover similarly to the spectral weight. This 
might give rise to a crossover in the spectral function as well. 

In summary the exponent $\beta$ in (\ref{beta}) characterizes the 
power-law singularity of the spectral function and the logarithmic decay of 
the quasiparticle weight. The exponent does not depend 
on the mass of the impurity except in the static limit $t_h=0$ and it has a
discontinuity at $t_h=0$. Notice that the perturbative results agree with the 
the small-$U$ expansions of the exponent for $t_h=0$ and $t_h=t$.

\section{Numerical study}
\label{numeri}

This section presents a numerical study of the exponent $\beta(t_h)$ 
based on a variational approach. The variational predictions for the 
energy and correlation functions are compared to results from Lanczos exact 
diagonalizations and Projection Quantum Monte Carlo simulations. Then the 
variational calculations are used to extract the exponent.
 
The variational wavefunction was originally proposed for the single spin-flip 
problem in the 2D-Hubbard model in reference to the stability of the 
ferromagnetic state \cite{edw}. It was also used to study numerically the 
quasiparticle weight in two dimensions \cite{sor1}. Furthermore, this 
variational approach is equivalent to the approximation used in Ref. \cite{per}.
In one dimension, the variational class of wavefunctions contains all the 
eigenstates of the model (\ref{H}) for $t_h=t$, as shown by Edwards \cite{edw}.
The variational approach is thus expected to include much of the relevant 
correlations even for $t_h\neq t$. 

In the reference frame comoving with the impurity a wavefunction 
$\mid\Psi_q\rangle$ of total momentum $q$ is represented by a function 
$f(j_1,\ldots,j_N)$ depending only on the positions of the electrons:
\begin{equation}
\mid\Psi_q\rangle=\frac{1}{\sqrt{L}}\sum_{j_0=1}^{L}e^{iqj_0} d^{\dagger}_{j_0}
\sum_{j_1,\ldots,j_N=1}^{L}f(j_1-j_0,\ldots,j_N-j_0)
c^{\dagger}_{j_1}\ldots c^{\dagger}_{j_N} \mid 0 \rangle.
\end{equation}
The variational ansatz for $f$ is a determinant of single-particle 
wavefunctions $\phi_m$:
\begin{equation}
f(j_1,\ldots,j_N)=\frac{1}{\sqrt{N!}}\det\left[\phi_m(j_l)\right]_
{m,l=1,\ldots,N}.
\label{BAwav}
\end{equation}
The expectation value of the energy is: 
\begin{eqnarray}
\langle\hat{H}\rangle=&\sum_{l=1}^{N}\left[-t\sum_{j=0}^{L-1}(\phi_l^{\ast}(j)
 \phi_l(j+1)+c.c.)+U\mid\phi_l(0)\mid^2\right]
\nonumber\\
& -t_h \left[\exp(-iq)\det({\bf S}) + c.c.\right].
\nonumber
\end{eqnarray}
\begin{equation}
{\bf S}_{m n}=\sum_{j=0}^{L-1}\phi_m^{\ast}(j+1)\phi_n(j).
\end{equation}
The variational parameters $\phi_m(j)$ are found by minimization of the 
energy using a steepest descent algorithm. 
If one chooses as starting $\phi_l(j)$ the exact solution for $t_h=t$, 
convergence is reached after a relatively small number of iterations even 
for hopping parameters very different from $t$. 

{\it Comparison of variational results to Lanczos exact diagonalizations and 
Projection Quantum Monte Carlo simulations.} Only small systems are 
accessible by exact diagonalization because the dimension of the 
Hilbert space increases very rapidly with the number of lattice sites. 
For the half-filled band, the biggest closed-shell system studied has 
$L=22$ whereas for $\rho=1/4$ it has $L=20$. For larger systems, Monte Carlo 
simulations are required. 

Projection Quantum Monte Carlo gives a statistical estimate of ground-state 
expectation values of observables $\hat{O}$ by a projection of a trial 
wavefunction $\mid\psi_T\rangle$ onto the groundstate with the operator 
$\exp(-\eta\hat{H})$ \cite{sor2}:
\begin{equation}
\langle\psi_0\mid\hat O\mid\psi_0\rangle=lim_{\eta\rightarrow\infty}
  \frac{\langle\psi_T\mid\exp(-\eta \hat{H})\hat O \exp(-\eta \hat{H})
  \mid\psi_T\rangle}
  {||\exp(-\eta H)|\psi_T\rangle||^2}.
\end{equation}
The algorithm used here closely follows Ref. \cite{loh}. 
The imaginary time evolution $\exp(-\eta\hat{H})$ is performed sequentially 
for small time intervals $\Delta\tau$ and a Trotter decomposition is used 
for the kinetic-energy and the interaction terms in the Hamiltonian. Eventually 
the two-body term is represented by discrete Hubbard-Stratonovitch fields 
that mediate the interaction. For the present calculations a rather 
large $\eta=15$ is necessary in order to converge the 
relevant correlation functions, and the results are extrapolated to 
$\Delta\tau\to 0$ using several values of the time interval. 

The energy is computed for the groundstate, which has zero total momentum.
The relevant quantity is the correlation energy 
$e_c=E_0-\tilde{E}_0-U\rho$ where $E_0$ and $\tilde{E}_0$ are the interacting 
and noninteracting energy, respectively. Notice that $e_c$ is of order $1$ 
while $E_0$ is of order $N$. Table \ref{tab1} gives $e_c$ for several  
system sizes at half and quarter filling and for 
an interaction equal to the bandwidth $U=4t$ and a hopping parameter 
$t_h=0.5t$. The variational ansatz is not exact for $t_h\neq t$ since 
the variational energy departs significantly from the exact energy. 
Nevertheless it remains always very close to the ground-state energy. 
The relative difference between the exact and the variational 
energy remains smaller than $0.1 \char'45$ even  
when extrapolating to the infinite-system limit with a 
$1/N$ scaling law. For the sake of comparison the correlation energy for 
the unrestricted Hartree-Fock solution is $-0.4380t$ at 
quarter filling, a value much larger than the variational 
energy $-0.6413t$. 

The $k=0$ component of the momentum distribution function of 
the impurity $n(k=0)$ has the same scaling law as $Z$ when the Anderson 
orthogonality catastrophe occurs. Both $Z$ and $n(k=0)$ are computed in order 
to test the relevance of the variational calculations to extract the 
exponent $\beta$. Figure \ref{three} shows the ratio of the exact momentum 
distribution $n(k=0)$, calculated by either exact diagonalization or 
Quantum Monte Carlo depending on the system size, to the variational estimate
$n(k=0)_{var}$ as a function of $1/N$ at quarter and half filling, for $U=4t$ 
and $t_h=0.5t$. If the variational result reproduced correctly the scaling 
behavior of $n(k=0)$ with $N$, the ratio $n(k=0)/n(k=0)_{var}$ should be 
constant as a function of $1/N$. At half filling, the ratio remains indeed 
always close to $1$ and the variational approach seems to correctly describe 
the scaling behavior. At quarter filling however, the comparison relies 
mostly on Monte Carlo simulations where the estimate of $n(k=0)$ has strong 
statistical fluctuations for large system sizes, large autocorrelation times 
having an essential contribution to the error bars. The comparison to the 
variational results is therefore delicate but still the ratio remains close 
to $1$ within the error bars. Finally exact-diagonalization results for $Z$ 
are presented in Fig. \ref{four} for $U=\infty$, $\rho=1/2$ and different 
$t_h$. The slope of $\ln Z$ as a function of $\ln N$ is an estimate of 
the exponent $\beta$ and both exact-diagonalization results and variational 
calculations are in good agreement. These last results illustrate the 
robustness of the variational wavefunction even in the strong-coupling 
regime. 

{\it Finite-size scaling analysis of the exponent.} 
The finite size corrections to the Anderson orthogonality catastrophe 
in (\ref{Zlaw}), which were derived from the perturbative analysis, are
used now to extract numerically the exponent. It should be stressed that 
these $1/N$ corrections contrast with the slowly decaying $\ln \ln N$ 
corrections expected for the paramagnetic phase of the 1-D Hubbard model 
\cite{sor3}.

The exponent is extracted from numerical data on finite systems 
by scaling the slope of $\ln Z$ as a function of $\ln N$: 
the spectral weights $Z_1$ and $Z_2$ are computed for systems with a number 
of fermions $N_1$ and $N_2$, respectively, at a fixed density and a fixed 
ratio $r=N_1/N_2$; from the perturbative analysis in (\ref{Zlaw}), the 
slopes $(\ln Z_1-\ln Z_2)/(\ln N_2-\ln N_1)$ have a polynomial expansion 
in $1/\bar{N}$, $\bar{N}$ being the mean number of electrons 
$\bar{N}=(N_1+N_2)/2$: 
\begin{equation} 
\frac{\ln Z_1-\ln Z_2}{\ln N_1-\ln N_2}=-\beta+ 
\alpha_1\frac{1-r^2}{2r\ln r}\frac{1}{\bar{N}}+
O\left(\frac{1}{\bar{N}^2}\right).
\end{equation}
The exponent $\beta$ is estimated by a numerical fit of the slopes as 
$\bar{N}\to\infty$. In practice, it is not possible to keep the ratio 
$N_1/N_2$ exactly fixed while increasing $\bar{N}$ but the number of 
electrons can be adjusted such that $N_1/N_2$ approaches a fixed value for 
large $\bar{N}$. All the results presented here have been obtained for 
$N_1/N_2\simeq 0.7$. 

The scaling procedure is tested both in the perturbative regime for 
different $t_h$ and at finite $U$ for $t_h=t$, where $\beta$ 
is known analytically. Figure \ref{five} illustrates the scaling procedure 
for the perturbative regime where $\ln Z$ is computed from (\ref{Zper}). 
The inset shows that the scaling procedure is well behaved even for very large 
$\bar{N}$. Furthermore, the scaling analysis is 
essential to determine the dependence of $\beta$ on $t_h$ since the 
finite-size corrections differ in both sign and magnitude for different $t_h$.
Note also that the non-linear terms in $1/\bar{N}$ become increasingly important
with decreasing $t_h$. The spectral weight $Z$ is computed numerically for 
$t_h=t$ and different $U$ using the Bethe's ansatz wavefunction (\ref{BAwav}) 
and the results are fitted with a third-order polynomial in $1/\bar{N}$. The
relative accuracy of the fitting procedure in extracting $\beta$ remains of 
the order of $10^{-4}$ even for a strong interaction $U=8t$, as illustrated 
in Fig. \ref{six}. Although the scaling behavior was derived in the 
perturbative regime, it seems to hold for any interaction strength $U$. 

For a finite $U$ and $t_h\neq t$, the exponent is not known and its 
evaluation relies on the variational approach and the finite-size scaling 
analysis. Figure \ref{seven} illustrates the scaling procedure in the 
strong-coupling regime $U=\infty$, for $\rho=1/2$ and different $t_h$. 
For $t_h>0.1t$, the exponent, which is obtained by an extrapolation of the 
data as $N\to\infty$, depends only slightly on the order of the polynomial 
used in the fit. For $t_h=0.1t$, however, the fitting procedure is not 
well-behaved and a relative error of a few percents is expected in the 
extraction of $\beta$. A precise investigation of the heavy-mass regime 
$t_h\ll t$ would require the study of even larger systems due to the 
important nonlinear corrections in $1/N$. 

{\it Mass dependence of the exponent.}
The study of $\beta(t_h)$ is based on the variational approach and the 
scaling analysis presented above. Figure \ref{one} presents the
exponent normalized to its value at $t_h=t$, $\beta(t_h)/\beta(t_h=t)$, 
as a function of $t_h/t$. For a finite $U$, the exponent does depend on 
the hopping parameter $t_h$, in contrast to the perturbative result
$\beta(t_h)/\beta(t_h=t)=1$ indicated by the dotted line. Furthermore,
this dependence increases with increasing $U$ and in the strong-coupling 
regime $U=\infty$ the exponent varies quasi-linearly with $t_h/t$. 
For small $t_h$ and at half filling, the exponent only slightly departs from 
the linear behavior. Furthermore, the exponent depends on the density since 
the data for $U=\infty$ and $\rho=1/2$ significantly differ from the exponents 
for $\rho=1/3$. This contrasts with the exact exponent for $t_h=0$ and 
$t_h=t$ which are independent of the density. However, the limiting value 
$\lim_{t_h\to 0}\beta(t_h)$ seems independent of the density.

The numerical results demonstrate the discontinuity of the exponent at 
$t_h=0$ for a finite $U$. The numerical data, indeed, do not 
extrapolate to the exact results for $t_h=0$ indicated by the filled 
symbols in Fig. \ref{one}. A precise extraction of the limiting 
value $\lim_{t_h\to 0}\beta(t_h)$, however,  would require the investigation 
of larger systems.

The occurrence of the nonanalyticity at $t_h=0$ is not surprising since the 
translational symmetry is broken at this point. Still the role played by 
the recoil of the impurity in the discontinuity of the exponent is 
not clear. The perturbation calculations indicate that the discontinuity at 
$t_h=0$ is due to the irrelevance of the backscattering 
processes whenever $t_h>0$; yet a simple argument can persuade us that 
this is true only in the small $U$ regime. Indeed let us assume 
first backscattering to be responsible for the nonanalyticity for 
all $U$. One can devise an effective model 
for the heavy-impurity limit ($t_h\ll t$) where the impurity is considered 
static but the interaction with the electrons is restricted 
to forward scattering~: 
\begin{equation}
H=-2t\sum_k \cos k~c^{\dagger}_k c_k 
    + \frac{U}{L} \sum_{kk'>0}   c^{\dagger}_k c_{k'}.
\end{equation}
In this picture, the only effect of the impurity recoil is the restriction to 
forward-scattering processes. Since the potential is static, the exponent is 
expressed in terms of phaseshifts. A calculation of the phaseshift gives
the same exponent as the result for $t_h=t$ in (\ref{BAz}). This is not 
compatible with the numerical results 
for $t_h\ll t$. Therefore forward-scattering processes alone 
cannot account for the discontinuity and backscattering has to be invoked. 

\section{Conclusions}

This study of the one-dimensional Anderson orthogonality catastrophe, 
combining analytical and numerical calculations, has focussed on the effect 
of the impurity recoil. 

The numerical study requires a finite-size scaling analysis since
the Anderson orthogonality catastrophe results from a logarithmic decay of 
the quasiparticle weight with the number of fermions. The present work 
shows, however, that a reliable numerical analysis of the quasiparticle 
renormalization can be achieved if a precise scaling hypothesis is 
established and large enough systems are accessed. 

Within perturbation theory, the infrared singularity that 
signals the orthogonality catastrophe occurs for an impurity band of any 
dispersion. Still, there is a discontinuity between the zero-bandwidth 
exponent and the finite-bandwidth exponent. This discontinuity is 
related to the impurity recoil which opens a gap in the spectrum of 
particle-hole excitations for backscattering of one electron from the 
Fermi momentum $k_F$ to $-k_F$. 

Outside of the perturbative regime, the numerical analysis
demonstrates the discontinuous behavior of the exponent 
and agrees at half filling with a study of the heavy-mass and 
strong-coupling regime in Ref. \cite{thi}. At third filling however, 
the numerical results differ from the analytical prediction.

The discontinuity of the exponent is an asymptotic result valid only 
in the limit of an infinite system. For a finite number of 
electrons and a heavy but finite-mass impurity the quasiparticle 
weight has the same logarithmic behavior as a static impurity 
up to a critical number of electrons $N_c$ where the former weight shows a 
crossover to the true asymptotic decay for a finite-mass impurity. 
Furthermore, $N_c$ diverges with increasing mass of the impurity. This 
crossover is also expected in the low-frequency behavior of the spectral 
function.

\acknowledgments
I would like to thank X. Zotos, R. Car and J. Wilkins for their great help 
and support, as well as T. Kopp and A. Rosch for useful discussions. 
This work was supported at the institute IRRMA by 
the Swiss National Science Foundation and the University of Geneva, and at the 
Ohio State University by a young investigator grant from the Swiss National
Science Foundation and by the DOE - Basic Energy Sciences, Division of 
Material Sciences.

\appendix
\section*{}

This appendix derives the scaling law (\ref{Zlaw}) from the perturbative 
expression of the spectral weight $Z$ and estimates the crossover size 
$N_c$. The calculations are presented for the half-filled band in details 
and the density dependence of $N_c$ is briefly discussed at the end. 

{\it Forward scattering processes.} An electron with momentum $k_1$ is 
scattered into an empty state with momentum $k_2$ such that $k_1k_2>0$. 
The infrared singularity is caused by excitations around the Fermi 
momentum $k_F=\pi/2$ whose energy vanishes linearly with the momentum transfer 
$k_1-k_2$. For $k_1$ and $k_2>0$, the momenta are written 
as $k_1=k_F-2\pi n_1/L$, $k_2=k_F+2\pi n_2/L$ and the excitation energy is:
\begin{equation}
\Delta\epsilon(k_1,k_2)\simeq 4\pi t\frac{n_1+n_2}{2N}
\left(1+\frac{t_h\pi (n_1+n_2)}{2Nt}\right).
\end{equation}
While the linearization of the energies allows an exact calculation of the 
exponent $\beta(t_h)$, it provides only an estimation of the remaining terms 
in (\ref{Zlaw}). The contribution of forward-scattering processes to 
the spectral weight is:
\begin{equation}
-\left(\frac{U}{L}\right)\sum_{k_1k_2>0}\frac{1}{\Delta\epsilon(k_1,k_2)^2}
=-\frac{1}{2}(UN_F)^2\sum_{n_1=0}^{N/2}\sum_{n_2=1}^{N/2}\frac{1}{(n_1+n_2)^2
(1+\pi t_h(n_1+n_2)/(2Nt))^2}.
\label{a1}
\end{equation}
The fraction in the sum is expanded in four different terms:
\begin{eqnarray}
\frac{1}{(n_1+n_2)^{2}(1+\pi t_h(n_1+n_2)/(2Nt))^2}=\frac{1}{(n_1+n_2)^2}
+\frac{1}{(n_1+n_2+2Nt/(\pi t_h))^2}\nonumber\\
-\frac{\pi t_h}{Nt(n_1+n_2)}+\frac{\pi t_h}{Nt(n_1+n_2+2Nt/(\pi t_h))}.
\label{a2}
\end{eqnarray}
The first term gives the logarithmic divergence \cite{table}:
\begin{equation}
\sum_{n_1=0}^{N/2}\sum_{n_2=1}^{N/2}\frac{1}{(n_1+n_2)^2}=
\ln N+1+C-2\ln 2+\frac{1}{N}+O\left(\frac{1}{N^2}\right).
\label{a3}
\end{equation}
where $C\simeq 0.5772$ is Euler's constant. The contributions of 
all other terms remain finite for $N\to\infty$. 
For the second and fourth terms in(\ref{a2}), the discrete sum can be
replaced by an integral. Further the finite-size corrections are of order 
$1/N$. The extraction of the finite-size corrections for the third 
term, however, requires the evaluation of the discrete sum:
\cite{table}:
\begin{equation}
\frac{1}{N}\sum_{n_1=0}^{N/2}\sum_{n_2=1}^{N/2}\frac{1}{n_1+n_2}=
\ln 2+\left(\ln 2-\frac{1}{2}\right)\frac{1}{N}+O\left(\frac{1}{N^2}\right).
\end{equation}

{\it Backscattering processes.} The initial and final momenta have a different 
sign, $k_1k_2<0$ and, for $k_1>0$ and $k_2<0$, the momenta are written as 
$k_1=k_F-2\pi n_1/L$ and $k_2=-k_F-2\pi n_2/L$. The linearized excitation 
energy is:
\begin{equation}
\Delta\epsilon(k_1,k_2,)=4 \left(t_h+\frac{t\pi (n_1+n_2)}{2N} \right).
\end{equation}
For $t_h=0$, backscattering processes are gapless excitations and have 
the same contribution to $Z$ as forward-scattering processes, as given in 
(\ref{a3}). For $t_h>0$, however, they have a gap and their contribution, 
which is finite, is computed by an integral representation of the sum. 

Finally, all the terms, for both forward-scattering and backscattering, 
give finite size corrections of order $1/N$. Furthermore the coefficient 
$\alpha_0$ in (\ref{Zlaw}) is evaluated explicitely in order to extract 
the crossover size $N_c$:
\begin{equation} 
\alpha_0(t_h=0)=-\beta(0)\left(1+C-2\ln 2\right).
\end{equation}
\begin{equation} 
\alpha_0(t_h>0)=-\beta(t_h)\left[1+C+\ln\frac{2t(\pi t_h+2t)}
{(\pi t_h+4t)^2}+\frac{\pi t_h}{t}\ln\frac{\pi t_h+2t}
{\pi t_h+4t}+\ln\frac{(\pi t+4t_h)^2}{8t_h(\pi t+2t_h)}\right].
\end{equation}
The crossover size $N_c$ is the solution of the equation: 
\begin{equation}
-\beta(0)\ln N_c +\alpha_0(0)=-\beta(t_h)\ln N_c+\alpha_0(t_h)
\end{equation}
The crossover occurs for $t_h\ll t$ where $\alpha_0(t_h)$ is diverging 
logarithmically. As a consequence $N_c$ is proportional to $t_h/t$:
\begin{equation}
N_c=\frac{\pi t}{2t_h}\exp(-1-C)\simeq 0.3244\frac{t}{t_h}.
\label{cross}
\end{equation}

{\it Arbitrary density.} The scaling law (\ref{Zlaw}) applies to any 
density. Furthemore the crossover size is proportional to  $t/t_h$ and 
diverges as $1-\rho$ in the limit $\rho\to 1$.

\begin{figure}[htb]

\centerline{\psfig{file=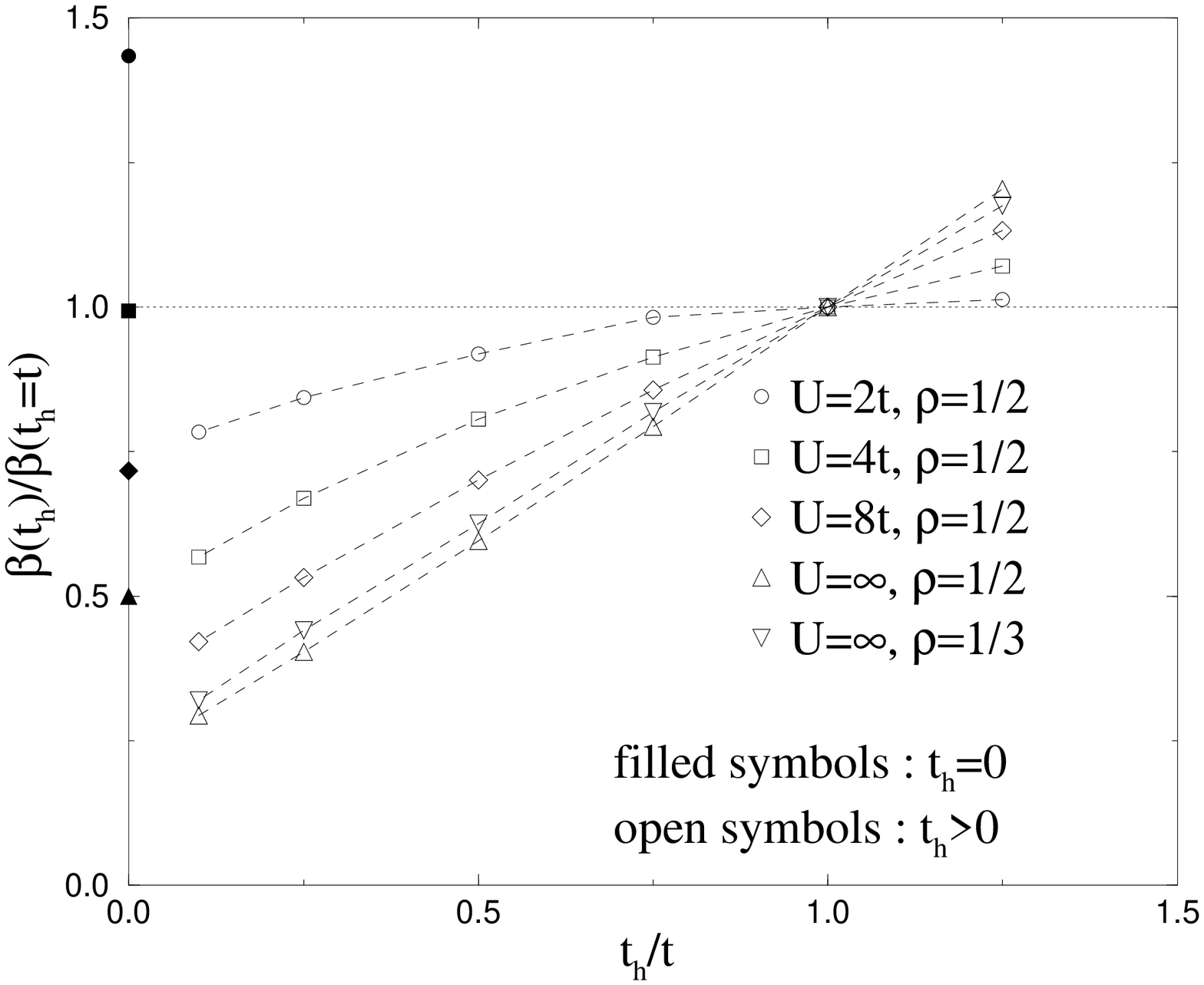 ,width=8cm,angle=-90}}
\caption{Exponent $\beta(t_h)/\beta(t_h=t)$ normalized to the analytical 
value $\beta(t_h=t)$ in (\ref{BAz}) as a function of the mass ratio 
$t_h/t$ for different interaction strengths $U$ and densities $\rho$. The 
filled symbols at $t_h=0$ are the exact results in(\ref{beta0}) for the 
static impurity. The open symbols for $t_h>0$ are the exponents extracted 
as $N\to\infty$ from the variational results for a sequence of finite systems
with $N_1/N_2=0.7$, the largest size being $N_1=159$ and $N_1=121$ at half and 
third filling, respectively.} 
\label{one}

\centerline{\psfig{file=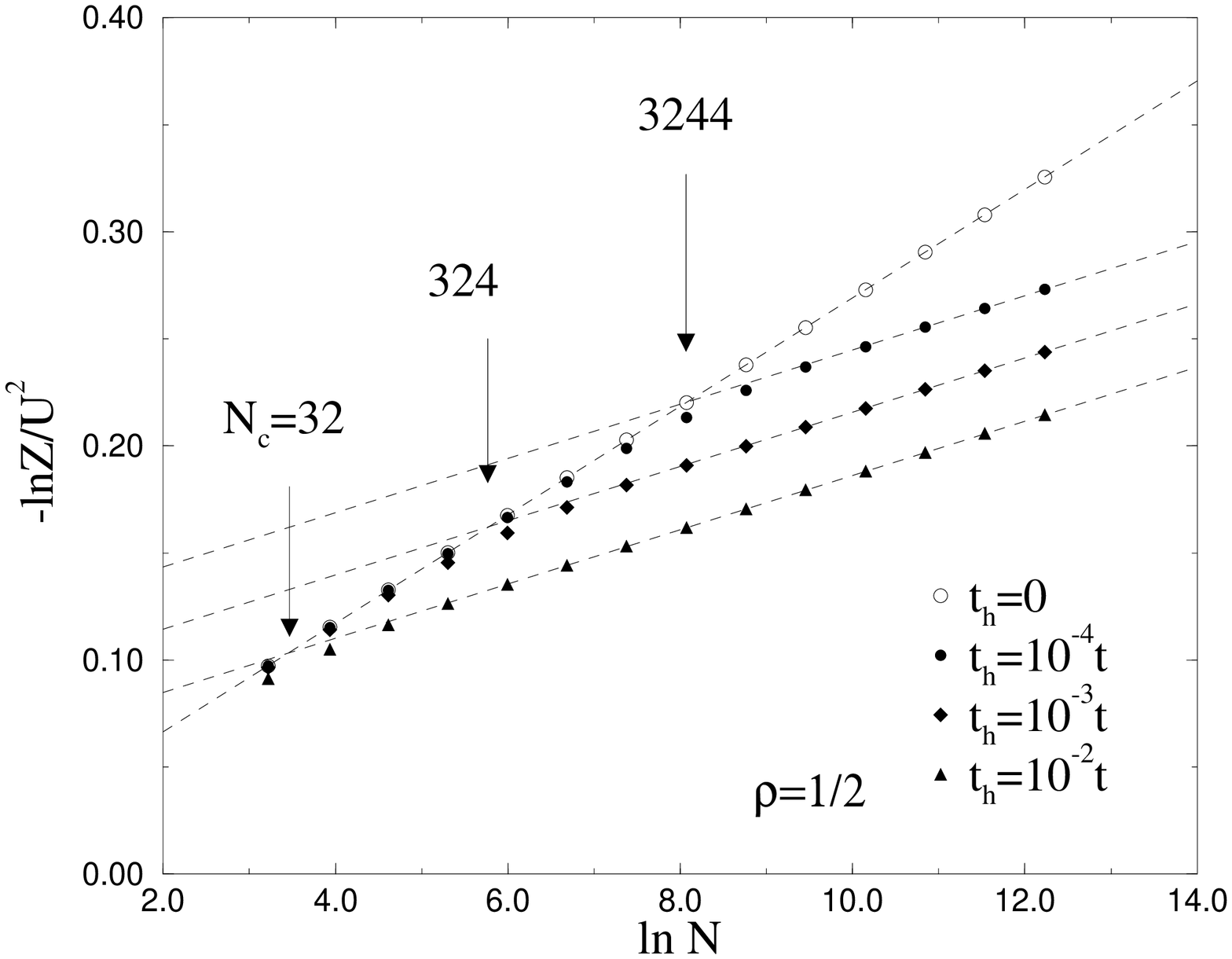 ,width=8cm,angle=-90}}
\caption{Logarithmic decay of the quasiparticle weight $Z$ with number of 
electrons $N$ from the perturbative result in (\ref{Zper}) at half filling 
and for different impurity hopping parameters $t_h$. The dashed lines are 
the asymptotic behaviors with the analytical exponent $\beta(t_h)$ in 
(\ref{beta}). The arrows indicate the estimate of the crossover size 
$N_c$ from (\ref{cross}).}
\label{two}

\centerline{\psfig{file=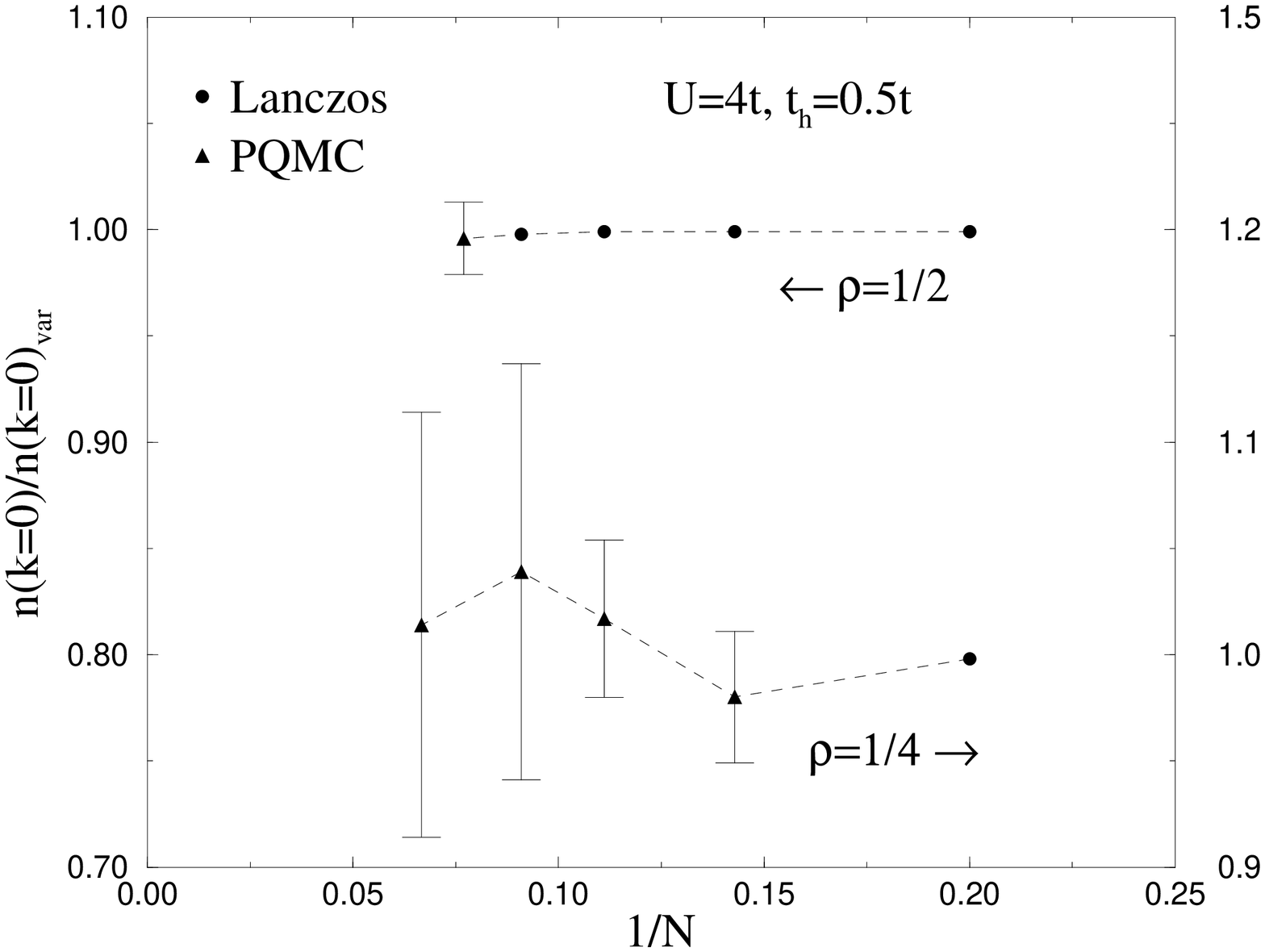 ,width=8cm,angle=-90}}
\caption{Ratio of the momentum distribution function $n(k)$ at $k=0$,
calculated with Lanczos diagonalization or Projection Quantum Monte Carlo,
to its variational estimate $n(k=0)_{var}$ as a function of $1/N$ 
for half and quarter filling and for $U=4t$, $t_h=0.5t$. The circles are the 
Lanczos results and the squares the Monte Carlo results with the corresponding 
error bars. The left axis refers to half filling and the right axis to 
quarter filling.}
\label{three}
 
\centerline{\psfig{file=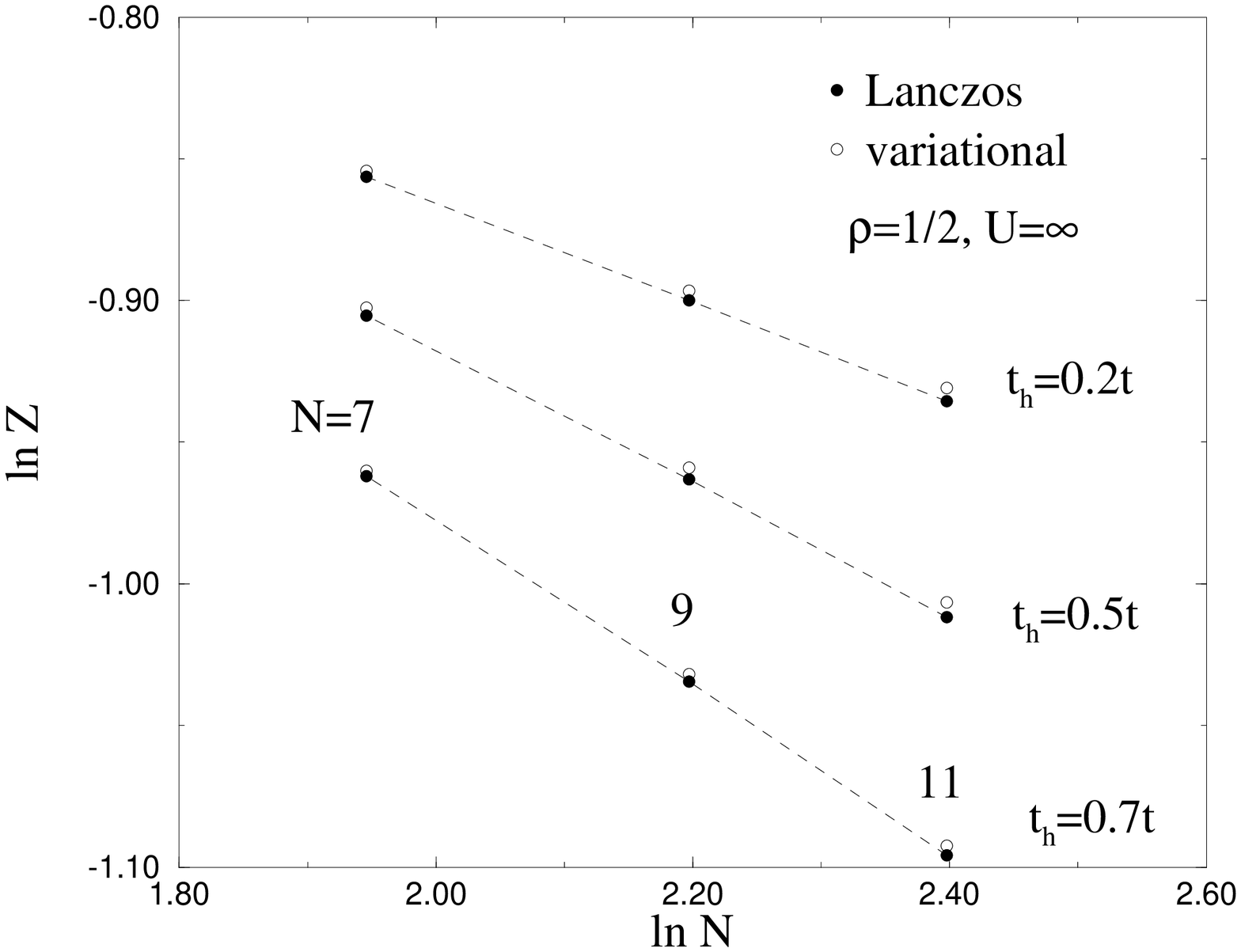 ,width=8cm,angle=-90}}
\caption{Comparison of exact spectral weight $Z$ from Lanczos diagonalization 
to its variational estimate as a function of number of electrons $N$ in log-log 
plot for $U=\infty$, $\rho=1/2$ and different $t_h$. The open symbols are the 
variational results and the filled symbols Lanczos results.}
\label{four}

\centerline{\psfig{file=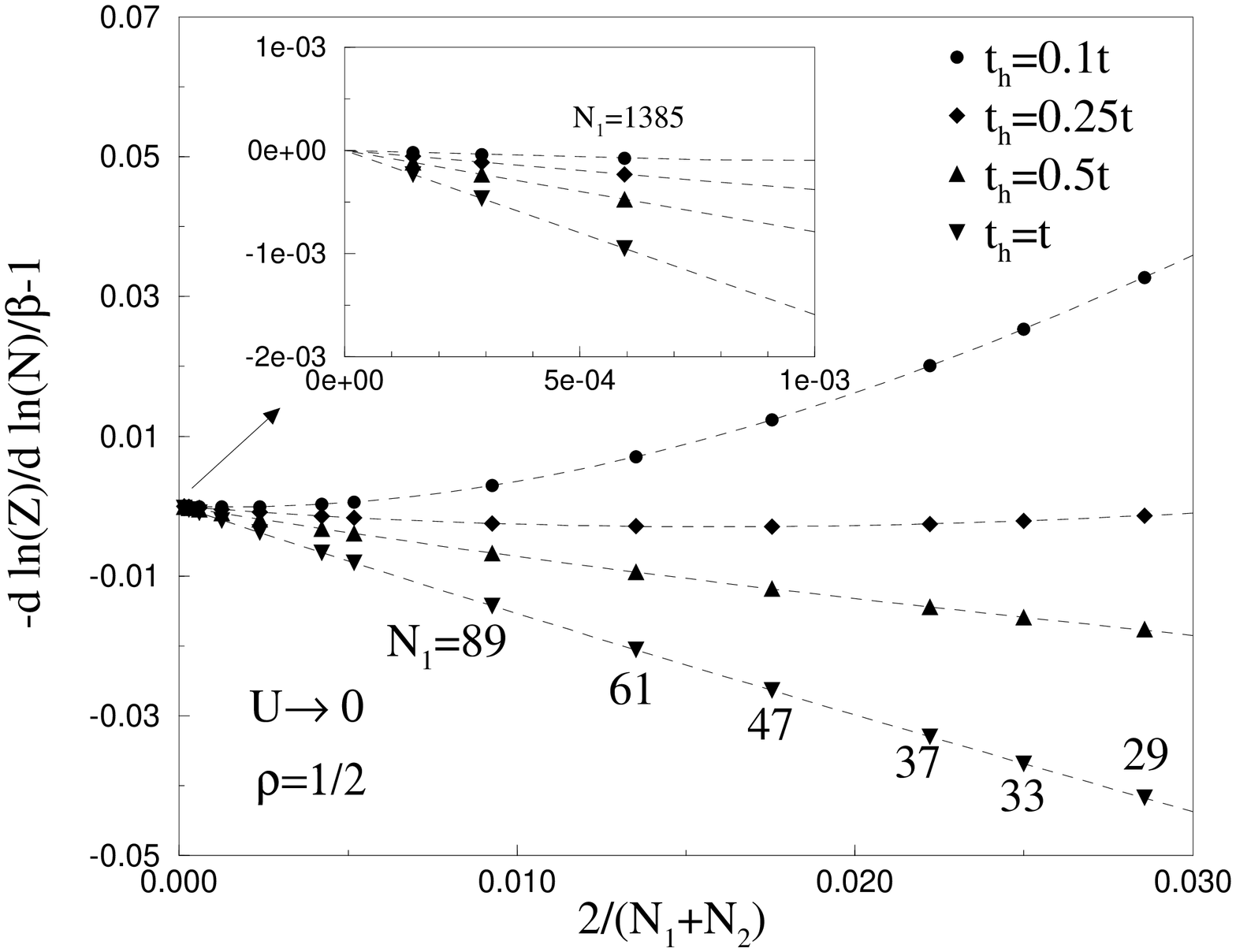 ,width=8cm,angle=-90}}
\caption{Slopes $d \ln(Z)/d \ln(N)=(\ln Z_1-\ln Z_2)/(\ln N_1-\ln N_2)$
normalized to the analytical value of $\beta(t_h)$ in (\ref{beta}) 
as a function of the inverse mean number of electrons $2/(N_1+N_2)$ in the 
perturbative regime and for different hopping parameters $t_h$. 
The spectral weight $Z$ is computed numerically from (\ref{Zper}). 
The dashed lines are numerical fits of the data as polynomials in 
$2/(N_1+N_2)$. The inset is a blow-up around the origin.}
\label{five}

\centerline{\psfig{file=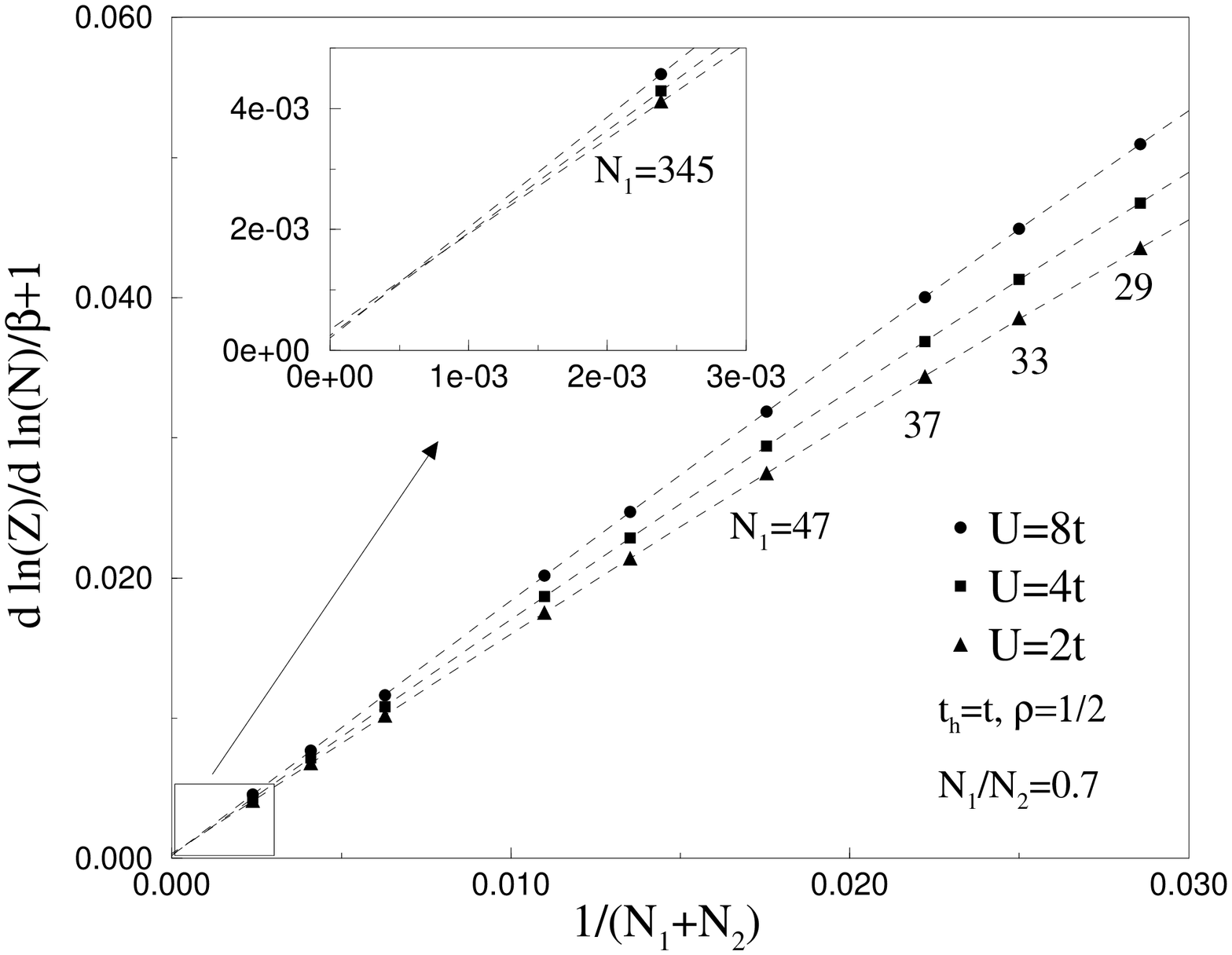 ,width=8cm,angle=-90}}
\caption{Slopes $d \ln(Z)/d \ln(N)$ normalized to the analytical value of 
$\beta(t_h=t)$ in (\ref{BAz}) as a function of the inverse mean number of 
electrons $2/(N_1+N_2)$ for $t_h=t$ and different interaction strengths 
$U=2,4,8t$ at half filling and for $N_1/N_2=0.7$. The spectral weight $Z$ is 
computed with the Bethe's ansatz wavefunction in (\ref{BAwav}). The dashed 
lines are numerical fits of the data as a third-order polynomial in 
$2/(N_1+N_2)$. The inset is a blow-up around the origin.}
\label{six}

\centerline{\psfig{file=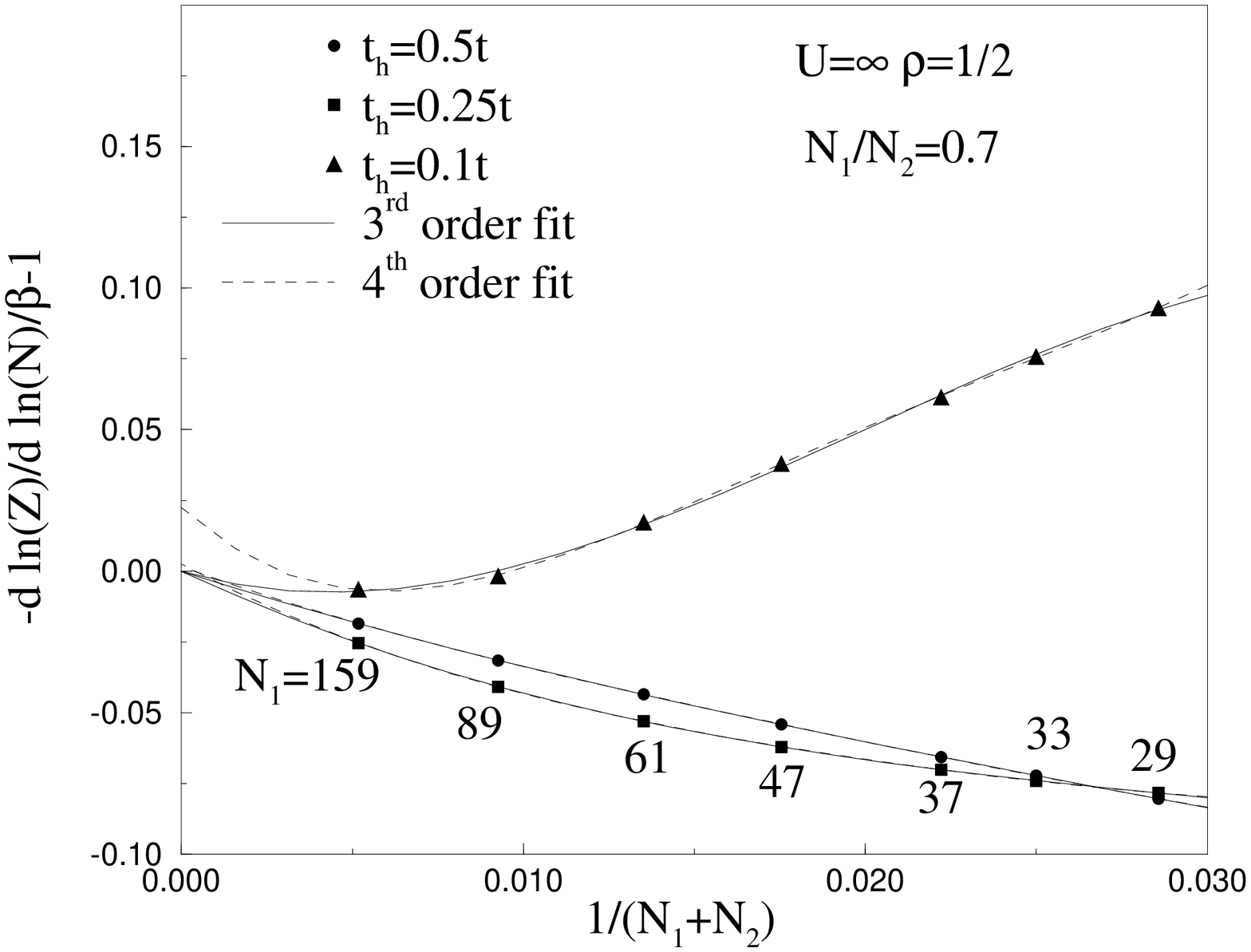 ,width=8cm,angle=-90}}
\caption{Slopes $d \ln(Z)/d \ln(N)$ normalized to the fitted $\beta(t_h)$ 
as a function of the inverse mean number of electrons $2/(N_1+N_2)$ for
$U=\infty$, different hopping parameters $t_h=0.5, 0.25$ and $0.1t$, at half 
filling. The spectral weight is computed with the variational wavefunction. 
The solid and dashed lines are numerical fits of the data as third-order and 
fourth-order polynomial in $2/(N_1+N_2)$, respectively.}
\label{seven}
\end{figure}

\begin{table}

\begin{tabular}{ccccccccc}
&&$e_c$&&&&&$e_c$&\\
 $N$\,($\rho$=1/2)&Lanczos&PQMC&variational&$\quad\quad$&
 $N$\,($\rho$=1/4)&Lanczos&PQMC&variational\\
\tableline
5&-0.9028&&-0.9025&&
5&-0.6102&&-0.6101\\
7&-0.9230&&-0.9227&&
7&&-0.619(4)&-0.6190\\
9&-0.9340&&-0.9336&&
9&&-0.624(3)&-0.6239\\
11&-0.9408&&-0.9404&&
11&&-0.627(2)&-0.6271\\
13&&-0.945(2)&-0.9451&&
15&&-0.632(2)&-0.6309\\
\tableline
$\infty$&-0.9727&&-0.9720&&
$\infty$&&-0.642(4)&-0.6413\\
\end{tabular}
\caption{Comparison of correlation energy, $e_c=E_0-\tilde{E}_0-U\rho$ 
where $E_0$ and $\tilde{E}_0$ are the ground-state energies of the 
interacting and noninteracting system, respectively, 
computed by Lanczos exact diagonalizations, Projection Quantum Monte Carlo 
(PQMC) and Edwards' variational wavefunction, as a function of the number 
of fermions $N$ at half and quarter filling ($\rho=N/L=1/2$ and $1/4$, 
respectively), for $t_h=0.5t$, $U=4t$. The last line gives 
the energy extrapolated for the infinite system both from either Lanczos 
or PQMC results and the variational results by a linear fit in $1/N$.}
\label {tab1}

\end{table}

\end{document}